\documentclass[twocolumn,10pt,aps,prl,nofootinbib]{revtex4-1}
\usepackage{amssymb, amsmath, amsthm}
\usepackage{graphicx}
\usepackage{units}

\newcommand{\ket}[1]{\ensuremath{\vert{#1}\rangle}}

\usepackage{color}
\usepackage[usenames,dvipsnames]{xcolor}
\usepackage{url}

\begin{document}
\title{Non-equilibrium phase transition in a dilute Rydberg ensemble}
\author{C. Carr}
\author{R. Ritter}
\author{C. G. Wade}
\author{C. S. Adams}
\author{K. J. Weatherill}
\affiliation{Joint Quantum Centre (JQC) Durham-Newcastle, Department of Physics, Durham University, South Road, Durham, DH1 3LE, United Kingdom}
\begin{abstract}
\noindent We demonstrate a non-equilibrium phase transition in a dilute thermal atomic gas. The phase transition, between states of low and high Rydberg occupancy, is induced by resonant dipole-dipole interactions between Rydberg atoms. The gas can be considered as dilute as the atoms are separated by distances much greater than the wavelength of the optical transitions used to excite them. In the frequency domain we observe a mean-field shift of the Rydberg state which results in intrinsic optical bistability above a critical Rydberg number density. In the time domain we observe critical slowing down where the recovery time to system perturbations diverges with critical exponent $\alpha = -0.53\pm0.10$. The atomic emission spectrum of the phase with high Rydberg occupancy provides evidence for a superradiant cascade.
\end{abstract}
\date{\today}
\maketitle

\noindent Non-equilibrium systems displaying phase transitions are found throughout nature and society, for example in ecosystems, financial markets and climate \cite{Hake80}. The steady state of a non-equilibrium system is a dynamical equilibrium between driving and dissipative processes. In atomic physics, one of the most studied non-equilibrium phase transitions is optical bistability where the driving is provided by a resonant laser field and the dissipation is inherent in the atom-light interaction. In most examples of optical bistability feedback is provided by an optical cavity, as in the pioneering work of Gibbs \cite{Gibb76,Gibb85}. However, bistability can also arise in systems where many dipoles are located within a volume which is much smaller than the optical wavelength; in this case the feedback is due to resonant dipole-dipole interactions \cite{Carm77, Wall78}. This latter case is known as intrinsic optical bistability \cite{Bowd79} and has, so far, only been observed in an up-conversion process between densely packed Yb$^{3+}$ ions in a solid-state crystal host cooled to cryogenic temperatures \cite{Hehl94}.  Intrinsic optical bistability generally cannot be observed for simple two-level systems such as atomic gases, because the resonance broadening, which is larger than the line shift \cite{Keav12}, suppresses the bistable response \cite{Hopf84, Frie89}.

A solution to this problem is provided by highly-excited Rydberg states, where the dipole-dipole induced level shifts between neighbouring states can be much larger than the excitation linewidth. This property of optical excitation of Rydberg atoms, known as dipole blockade \cite{Luki01}, enables a diverse range of applications in quantum many-body physics, quantum information processing \cite{Saff10}, non-linear optics \cite{Prit13} and quantum optics \cite{Dudi12, Peyr12, Maxw13,Scha12}. An interesting feature of Rydberg systems is that the range of the interaction can be much larger than the optical excitation wavelength, giving rise to non-local interactions \cite{Sevi11}. This also creates the possibility of observing intrinsic optical bistability, and hence non-equilibrium phase transitions \cite{Lee11} over macroscopic, optically-resolvable length scales.

\begin{figure}[t]
\includegraphics[width=3.3in]{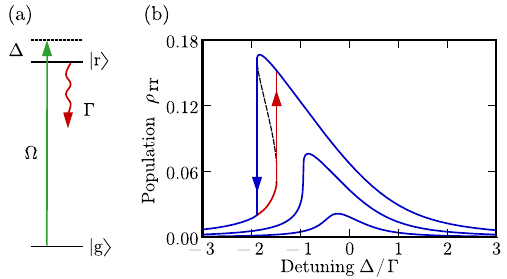}
\caption[]{(color online)
Theoretical model for the cooperative optical response of the ensemble. (a) Simplified two-level model with ground state $\ket{\rm g}$ and Rydberg state $\ket{\rm r}$. The levels are coupled by a laser with Rabi frequency $\Omega$ and detuning from resonance $\Delta$. (b) Rydberg state population $\rho_{\rm rr}$ as a function of laser detuning $\Delta$ for increasing Rabi frequency $\Omega$. As a result of the cooperative excitation-dependent shift, the response exhibits intrinsic optical bistability with hysteresis dependent upon the history of the ensemble. Theoretical parameters: interaction strength $V/\Gamma=-11$ and Rabi frequency $\Omega/\Gamma=(0.15,0.3,0.5)$.}
\end{figure}

\begin{figure}[t]
\begin{center}
\includegraphics[width=3.3in]{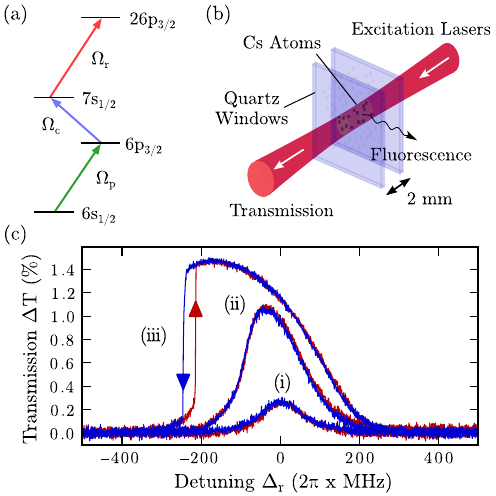}
\caption[]{(color online)
(a) Three-photon excitation scheme to Rydberg states in cesium. (b) Schematic of the experimental setup. The three excitation lasers co-propagate through a 2~mm vapour cell. The non-equilibrium dynamics are probed by measuring the transmission of the probe laser or analysing the emitted fluorescence. (c) Experimental optical response $\Delta T$ as a function of Rydberg laser detuning $\Delta_{\rm r}$ for Rydberg Rabi frequency $\Omega_{\rm r}$ increasing from (i) to (iii). Experimental parameters: ground state density ${\cal N}=4.3 \times 10^{12}$~cm$^{-3}$, probe Rabi frequency $\Omega_{\rm p}=2\pi\times37$~MHz, coupling Rabi frequency $\Omega_{\rm c}=2\pi\times77$~MHz and Rydberg Rabi frequency $\Omega_{\rm r}=2\pi\times$~(14,36,74)~MHz.}
\end{center}
\end{figure}

In this letter, we demonstrate a non-equilibrium phase transition in a thermal Rydberg ensemble. In contrast to previous experiments, we directly observe optical bistability in the transmission of the probe light without the requirement for cryogenics \cite{Hehl94} or cavity feedback \cite{Gibb76}. We distinguish between the phases of low and high Rydberg occupancy using fluorescence spectroscopy and confirm the first-order phase transition through the observation of critical slowing down in the temporal response of the ensemble.
Our observation of a non-equilibrium phase transition in a dilute atomic system provides a new platform to study the transition between classical mean field and microscopic quantum dynamics \cite{Lee12, Ates12}.

To illustrate the origin of the non-equilibrium phase transition in our system, we begin by considering the simple two-level atom shown in Fig.~1(a) with ground state $\ket{\rm g}$ and Rydberg state $\ket{\rm r}$. 
The levels are coupled by a laser with Rabi frequency $\Omega$ and detuning from resonance $\Delta$. 
Using standard semi-classical analysis, the time evolution of the system can be described using the Lindblad master equation for the single-particle density matrix. 
To include the effect of dipole-dipole interactions between atoms in state $\ket{\rm r}$,  we use a classical approximation to the many-body quantum dynamics \cite{Lee12} and introduce a mean-field shift of the Rydberg state, proportional to its steady state population. 
Coherent dynamics have been observed in thermal ensembles by operating on ultrashort time scales, where the Rydberg interactions appear as a dephasing of the Rabi oscillations due to the broad distribution of inter-atomic distances \cite{Balu13}. 
However, over longer time scales or in steady state as in our experiment, the interactions between Rydberg atoms result in a mean-field shift. In this limit, the laser detuning $\Delta\rightarrow\Delta-V\rho_{\rm rr}$ where $V$ is the dipole-dipole interaction term and $\rho_{\rm rr}$ is the population of the Rydberg state. 
The interaction term $V$ corresponds to the sum of the dipole-dipole interaction over the excitation volume \cite{Frie73}.




The optical Bloch equations for the two-level system can be written as
\begin{subequations}
\begin{eqnarray}
\dot\rho_{\rm gr} &=& i\Omega \left( \rho_{\rm rr} - \frac{1}{2} \right) + i(\Delta-V\rho_{\rm rr})\rho_{\rm gr} - \frac{\Gamma}{2}\rho_{\rm gr} \\
\dot\rho_{\rm rr} &=& -\Omega~\text{Im}(\rho_{\rm gr}) - \Gamma\rho_{\rm rr}
\end{eqnarray}
\end{subequations}
where the off-diagonal coherence terms $\rho_{\rm rg}=\rho_{\rm gr}^*$ and the diagonal population terms $\rho_{\rm gg}=1-\rho_{\rm rr}$. 
The steady-state solution for the Rydberg population $\rho_{\rm rr}$ as a function of laser detuning $\Delta$ is shown in Fig.~1(b) for increasing Rabi frequency $\Omega$. As the cooperative shift is dependent on the Rydberg population, the lineshape becomes asymmetrical and eventually exhibits bistability with hysteresis dependent on the direction in which the detuning is varied (shown by the arrows). In the bistable region, there is also an unstable state (shown by the dashed curve) which cannot be measured experimentally. At the critical transition, there is an abrupt change in the atomic dynamics. The solution of Equation (1) can be found in the Supplementary Material \cite{supmat1}.

\begin{figure*}[t]
\begin{center}
\includegraphics[width=6.2in]{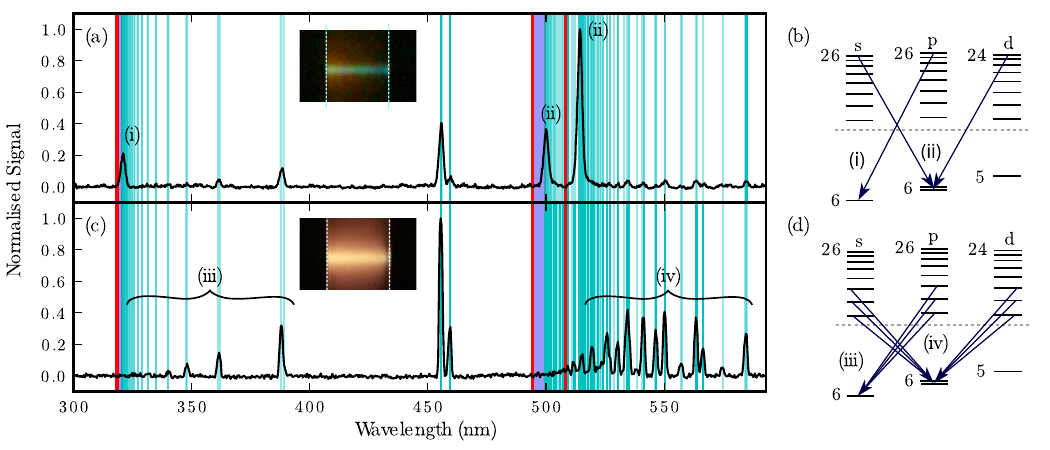}
\caption[]{(color online)
Atomic emission spectra for low and high Rydberg state occupancy. The visible fluorescence spectrum is shown for (a) ${\cal N}=3.1 \times 10^{11}$~cm$^{-3}$ and (c) ${\cal N}=4.3 \times 10^{12}$~cm$^{-3}$. In the low Rydberg occupancy phase, the spontaneous emission originates from high-lying Rydberg states as illustrated in (b). However, in the high Rydberg occupancy phase, the spontaneous emission originates from low-lying Rydberg states, as illustrated in (d), due to a superradiant cascade between high-lying Rydberg states. The ionisation limits from 6s$_{1/2}$, 6p$_{1/2}$ and 6p$_{3/2}$ are shown by thick red vertical lines. The blue shaded regions highlight the absence of spontaneous emission between 26p$_{3/2}$ and ionisation which would occur due to a blackbody or collisional excitation process. The thin cyan vertical lines indicate the dipole-allowed transitions. Probe Rabi frequency $\Omega_{\rm p}=2\pi\times41$~MHz, coupling Rabi frequency $\Omega_{\rm c}=2\pi\times74$~MHz and Rydberg Rabi frequency $\Omega_{\rm r}=2\pi\times122$~MHz.}
\end{center}
\end{figure*}

To experimentally observe a non-equilibrium phase transition in a dilute medium we use a resonant multi-photon excitation scheme in a thermal Cs vapour, as shown in Fig. 2(a) \cite{Carr12a}. In the simple theoretical analysis above Doppler averaging is not considered; however by using a multi-photon scheme we excite only a narrow velocity distribution of atoms \cite{tana12} and can therefore access a regime where the mean-field shift between Rydberg states far exceeds the Doppler width of the excitation.  The optical Bloch model for the  multi-photon scheme is presented in the Supplementary Material \cite {supmat2}. A schematic of the experimental setup is shown in Fig.~2(b). A thermal vapour of Cs atoms is confined in a quartz cell with an optical path length of 2~mm. The atoms are driven into the 26p$_{3/2}$ Rydberg state using three excitation lasers which co-propagate through the cell. The probe laser, with wavelength $\lambda_{\rm p}=852.3$~nm, Rabi frequency $\Omega_{\rm p}$ and waist $w_{\rm p}=150$~$\mu$m is frequency stabilised to the ${| 6s_{1/2},F=4 \rangle \rightarrow | 6p_{3/2},F^\prime=5\rangle}$ transition. The coupling laser, with wavelength $\lambda_{\rm c}=1469.9$~nm, Rabi frequency $\Omega_{\rm c}$ and waist $w_{\rm c}=80$~$\mu$m is stabilised to the ${| 6p_{3/2},F^\prime=5 \rangle \rightarrow | 7s_{1/2},F^{\prime\prime}=4\rangle}$ transition using excited state polarisation spectroscopy \cite{Carr12} . Finally, the Rydberg laser with wavelength $\lambda_{\rm r}=790.3$~nm, Rabi frequency $\Omega_{\rm r}$ and waist $w_{\rm r}=80$~$\mu$m, is tuned around the resonance between the excited-state 7s$_{1/2}$ and the Rydberg state 26p$_{3/2}$.

For a multi-photon transition to a Rydberg state, the transmission of the probe light resonant with the optical transition is increased by population shelving in the Rydberg state \cite{Thou09} and provides a direct readout of the Rydberg population. The change in probe laser transmission $\Delta T$ as a function of Rydberg laser detuning $\Delta_{\rm r}$ is shown in Fig.~2(c) for increasing Rydberg Rabi frequency $\Omega_{\rm r}$. As the level of Rydberg population increases, the excitation-dependent shift first produces an asymmetry in the lineshape (ii). Eventually, when the shift is greater than the linewidth (iii), the lineshape exhibits intrinsic optical bistability with hysteresis dependent on the direction in which resonance is approached (shown by the arrows). Importantly, this bistability is measured in steady-state and is not a transient phenomenon. As a result, within the hysteresis window, the system can be placed in either the low or high Rydberg occupancy phase with exactly the same experimental parameters. 

\begin{figure}[t]
\begin{center}
\includegraphics[width=3in]{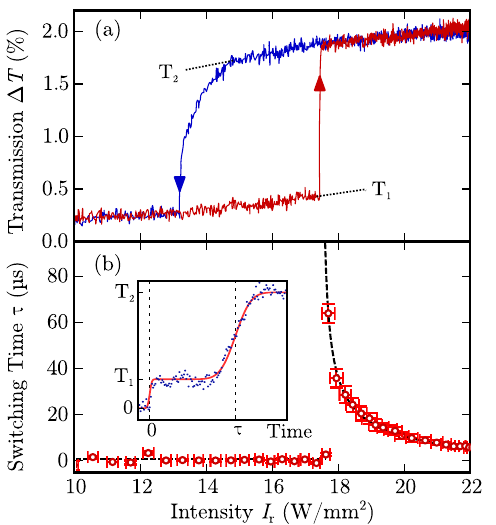}
\caption[]{(color online)
Critical slowing down as the temporal signature of a phase transition. (a) Continuous Rydberg laser intensity $I_{\rm r}$ scan showing bistability and hysteresis in the optical response $\Delta T$. (b) Discrete Rydberg laser intensity $I_{\rm r}$ scan showing the divergence of the switching time to steady-state $\tau$ around the critical transition intensity ${I}_{\rm r,crit}\approx17.5$~W/mm$^2$. The switching time diverges as $({{I}_{\rm r}-{I}_{\rm r,crit}})^{\alpha}$ with critical exponent $\alpha=-0.53\pm0.10$ (standard deviation error) shown by the dashed line of best fit. Ground state density ${\cal N}=4.3 \times 10^{12}$~cm$^{-3}$, probe Rabi frequency $\Omega_{\rm p}=2\pi\times57$~MHz, coupling Rabi frequency $\Omega_{\rm c}=2\pi\times~116$~MHz and Rydberg detuning $\Delta_{\rm r}=2\pi\times-220$~MHz. The error bars represent the standard deviation error on the determination of the laser intensity and switching time.}
\end{center}
\end{figure}

The change in atomic behaviour across the phase transition can be analysed by measuring the spectrum of the off-axis fluorescence. The emission spectra for the two phases of low and high occupancy are shown in Fig.~3(a) and (c), respectively. In the low phase, the dominant transitions indicated by (i) and (ii) involve decay from high-lying Rydberg states to the ground states of the s, p and d series. This behaviour, highlighted in Fig.~3(b), is consistent with spontaneous emission where such transitions dominate due to the $\omega^3$ dependence in the Einstein A-coefficient. This phase is characterised by the faint (green) fluorescence shown in the inset.

In the high Rydberg occupancy phase the emission spectrum is dramatically modified. The dominant spontaneous emission transitions (i) and (ii) are no longer present. Instead, the spontaneous emission now originates from a range of low-lying Rydberg states indicated by (iii) and (iv) and highlighted in Fig.~3(d). This phase is characterised by the strong (orange) fluorescence shown in the inset. Importantly, the absence of emission close to the ionisation limit for each series, indicated by the dark (red) vertical lines, indicates that atoms are not promoted to higher-lying Rydberg states, as would occur in a collisional or up-conversion processes \cite{Hehl94}. We can also neglect the effects of thermal blackbody photons because the average number of photons per mode at Rydberg-Rydberg transition frequencies is much lower than the average number of excited atoms \cite{Raim82}. The transitions at 455~nm and 459~nm occur in both phases and correspond to decay to 6s$_{1/2}$ from 7p$_{3/2}$ and 7p$_{1/2}$ respectively.

The emission spectrum in the high occupancy phase can be understood as a superradiant cascade to lower-lying Rydberg states \cite{Goun79}. Evidence for a superradiant cascade has also been observed in ultracold atoms \cite{Wang07,Weat08}. When the cooperativity on a particular transition is high, the atoms emit collectively and in-phase with one another. The single atom lifetime of the 26p$_{3/2}$ to 26s$_{1/2}$ transition is $\tau\simeq500$~$\mu$s but within the transition wavelength volume $V=1$~mm$^3$, we estimate the Rydberg atom number $N_{\rm r}\simeq5\times10^6$. Consequently, we expect a superradiant decay timescale of $\tau_{\rm super}=\tau/N_{\rm r}\simeq$~100~ps. As the transition wavelength and therefore the cooperative enhancement of the decay rate is proportional to $n^3$, the superradiant cascade eventually stops and gives rise to the observed spontaneous emission from low-lying Rydberg states as indicated in Fig.~3(iii) and (iv). 

By stabilising the laser frequency within the hysteresis window and varying the intensity of the Rydberg laser $I_{\rm r}$, it is possible to observe bistability and hysteresis in the optical response as shown in Fig.~3(a). The system switches between the low occupancy phase with probe transmission level $T_1$ and the high occupancy phase with probe transmission level $T_2$. In this case, the phase transition from low to high Rydberg population occurs at critical intensity ${I}_{\rm r,crit}$~$\approx$~17.5~W/mm$^2$.

The first-order phase transition between the low and high Rydberg occupancy phases can be confirmed through the observation of critical slowing down. This temporal signature of a phase transition occurs as the system approaches a critical point and becomes increasingly slow at recovering from perturbations \cite{Boni79,Sche09}. The temporal response of the ensemble is measured by discretely varying the Rydberg laser intensity ${I}_{\rm r}$ and measuring the time $\tau$ to reach steady-state, as illustrated in the inset of Fig.~3(b). At the critical transition, the switching time diverges according to the power law $\tau\propto({I}_{\rm r}-{I}_{\rm r,crit})^{\alpha}$ shown by the fitted dashed line. The critical exponent $\alpha=-0.53\pm0.10$ (standard deviation error) is consistent with previous work on first-order phase transitions and optical bistability \cite{Gryn83,Hohe77}.

We also note that the geometry of the excitation region plays an important role in our observation of many-body dynamics \cite{Frie11}. The optical path length of 2~mm provided by the vapour cell is comparable to the interaction wavelength. If the medium was much shorter, the cooperative shift would not result in intrinsic optical bistability. Furthermore, if the medium was much longer, the dipoles would not evolve with the same phase. A more complete study of the length dependence of the effect will form the focus of future work.

In summary, we have demonstrated a cooperative non-equilibrium phase transition in a dilute thermal atomic gas. The observations that have been discussed raise interesting possibilities for future non-local propagation experiments which utilise the long range cooperative interaction \cite{Sevi11}. Furthermore, this work could be used to perform precision sensing \cite{Abel11} around the critical point and to study resonant energy transfer \cite{Saro10} on optically-resolvable length scales. In addition, studies of the fluorescence in the vicinity of the phase transition could provide further insight into the dynamics of strongly-interacting dissipative quantum systems \cite{Ates12,Lee12}.

\begin{acknowledgements}
We would like to thank S A Gardiner and U M Krohn for stimulating discussions, R Sharples for the loan of equipment and M P A Jones and I G Hughes for proofreading the manuscript. CSA and KJW acknowledge financial support from EPSRC and Durham University. CSA and RR acknowledge funding through the EU Marie Curie ITN COHERENCE Network.
\end{acknowledgements}

\newpage
\begin{widetext}
\section{Non-equilibrium phase transition in a dilute Rydberg ensemble \\ \emph{Supplementary Material}}
\end{widetext}

In the main text, an analytic two level model is used to give an intuitive understanding of the system. In this supplementary material, we present:

\begin{enumerate}
\item Details of the analytic two level model.
\item A numerical solution of the full four level system.
\end{enumerate}

\noindent We justify the simplified 2-level system as an analogy for the 4-level system by presenting qualitatively similar predictions from both models.
\section{Analytic 2-Level Model}
\label{sec:2l}

The caesium vapor is described using optical Bloch equations. To model the nonlinear behavior we adjust the energy of the Rydberg state proportional to its population (Equation~\ref{eq:term}). This is a mean field approximation.
\begin{equation}
\label{eq:term}
\Delta \rightarrow \Delta - V\rho_{\mathrm{rr}}
\end{equation}

\noindent For a 2-level system, we start from the well known steady state solution of the 2-level optical bloch equations (\emph{see eg. Foot {\it Atomic Physics} (Oxford University Press, 2005}), and substitute the expression given in Eq.~\ref{eq:term}.
\vspace{-0.3cm}
\begin{equation}
\label{eq:standard}
\rho_{\mathrm{rr}} = \frac{\Omega^2/4}{(\Delta - V\rho_{\mathrm{rr}})^{2} + \Omega^2/2 + \Gamma^2/4}
\end {equation}

\noindent Rearranging, we write a cubic equation in $\rho_{rr}$ (Eq.~\ref{eq:cubic}). For each laser detuning, we find either one (monostable) or two (bistable) stable real valued solutions. In the bistable region, we also find an intermediate, unstable solution which we do not expect to realise experimentally.

\begin{equation}
\label{eq:cubic}
V^2\rho_{\mathrm{rr}}^3 - 2V\Delta\rho_{\mathrm{rr}}^2 + (\Delta^2 + \Omega^2/2 + \Gamma^2/4)\rho_{\mathrm{rr}} - \Omega^2/4 = 0
\end{equation}

\noindent This follows the work published by T.\ Lee {\it et al}. [20] very closely. The predicted bistable response of 2-level atomic vapor is shown in Fig.~1 of the main text and closely resembles Fig.~1 of this document

\section{Numerical 4-Level Model}
\label{sec:4l}

In the four level system with the probe and coupling lasers on resonance, the complete Hamiltonian, $H$, is given in a rotating frame as

\begin{equation}
\label{eq:mat}
H = \frac{\hbar}{2} \begin{pmatrix} 0 & \Omega_{\mathrm{p}} & 0 & 0 \\ 
\Omega_{\mathrm{p}} & 0 & \Omega_{\mathrm{c}} & 0 \\ 
0 & \Omega_{\mathrm{c}} & 0  & \Omega_{\mathrm{r}} \\ 
0 & 0 & \Omega_{\mathrm{r}} & - 2(\Delta_{\mathrm{r}} - V\rho_{\mathrm{rr}}) \end{pmatrix}
\end{equation}

To find steady state solutions of the optical bloch equations, each data point is calculated by evolving the Liouville Equation for the density matrix, $\rho$, (Eq.~\ref{eq:Lou}) in time until a steady state is reached. The matrix $\gamma$ implements phenomenological spontaneous decay between the atomic states.
\begin{equation}
\label{eq:Lou}
\dot{\rho} = \frac{i}{\hbar}\begin{bmatrix}\rho,H\end{bmatrix} - \gamma
\end{equation}

\begin{figure}
	\label{fig:neat}
  
  \begin{centering}
    \includegraphics[width=3.0in]{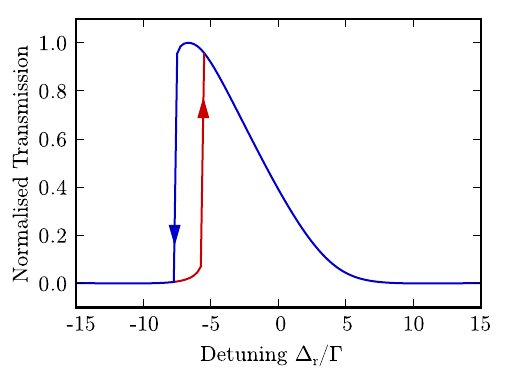}
		\caption{Normalised transmission of the probe laser with Rydberg laser detuning $\Delta_{\rm r}$ calculated using the four level optical Bloch model.}
\end{centering}
\end{figure}

In the bistable region where there are two steady states, both solutions are attained by using starting conditions close to one or other of the steady states. To simulate the intrinsic optical bistability reported in this letter, parameters as follows were used.
\begin{equation}
\label{eq:delta}
\begin{array}[b]{c}
\Omega_{\mathrm{p}} = 2\pi \times 110\ $MHz$ \\
\Omega_{\mathrm{c}} = 2\pi \times 200\ $MHz$ \\
\Omega_{\mathrm{r}} = 2\pi \times 30\ $MHz$ \\
V = 2\pi \times (-76.6)\ $MHz$\\
\end{array}
\end{equation}

\noindent Following the velocity class $v=0$, a bistable region is predicted as shown in Fig.~1. The transmission is normalised so that the maximum is 1 and the minimum 0.

Although we have shown qualitatively that both the 2-level and 4-level models predict bistability, we are aware of further complicating factors. Principally, the range of velocity classes present in the vapor will broaden the absorption line shape. Indeed, it is this effect that we suggest gives us the near Gaussian features in our experiment compared to the lorentzian shapes presented in these theoretical models. In addition the mean-field shift of -76.6~MHz used in the model is smaller than that required to observe optical bistability in the Doppler broadened resonance observed in the experiment.

\end{document}